\documentclass[%
 reprint,
showpacs,
 amsmath,amssymb,
 aps,pre,
]{revtex4-1}

\usepackage{graphicx}
\usepackage{dcolumn}
\usepackage{bm}
\usepackage{color}
\usepackage{epstopdf}
\usepackage{array,booktabs}
\usepackage{pdfpages}
\usepackage[colorlinks=true, citecolor=blue, filecolor=blue,urlcolor=blue, linkcolor=blue]{hyperref}

\usepackage{epsfig}

\newcolumntype{d}[1]{D{.}{.}{#1}}

\makeatletter
\newcommand{\thickhline}{%
    \noalign {\ifnum 0=`}\fi \hrule height 1pt
    \futurelet \reserved@a \@xhline
}
\newcolumntype{"}{@{\hskip\tabcolsep\vrule width 1pt\hskip\tabcolsep}}

\newcommand*{\colspacing}{\hskip \col@sep}

\newcolumntype{L}{>{\centering\arraybackslash}p{2.1cm}}
\newcolumntype{M}{>{\centering\arraybackslash}m{2cm}}
\newcolumntype{N}{>{\centering\arraybackslash}m{1.75cm}}
\newcolumntype{O}{>{\centering\arraybackslash}m{1.5cm}}
\makeatother
\begin{document}

\title{Coupled self-organization: \\ Thermal interaction between two liquid films undergoing long-wavelength instabilities}

\author{Mikl\'os V\'ecsei}
\affiliation{Center of Smart Interfaces, TU Darmstadt, Darmstadt, Germany}
\author{Mathias Dietzel}
\altaffiliation[Corresponding author: ]{dietzel@csi.tu-darmstadt.de}
\affiliation{Center of Smart Interfaces, TU Darmstadt, Darmstadt, Germany}
\author{Steffen Hardt}
\affiliation{Center of Smart Interfaces, TU Darmstadt, Darmstadt, Germany}


\begin{abstract}
The effects of thermal coupling between two thin liquid layers, separated by a gas layer, are discussed. The liquid layers undergo long-wavelength instabilities driven by gravitational and thermocapillary stresses. To study the dynamics, both a linear stability analysis and a full numerical solution of the thin-film equations are performed. The results demonstrate that the stability properties of the combined system differ substantially from the case where both layers evolve independently from each other. Most prominently, oscillatory instabilities, not present in single-liquid layer configurations, may occur.
\end{abstract}

\pacs{89.75.Fb, 47.54.-r, 47.20.Bp, 47.20.Ma \hspace{59pt}DOI: \href{http://dx.doi.org/10.1103/PhysRevE.89.053018}{10.1103/PhysRevE.89.053018} \\Published under the author rights of  the APS copyright agreement}

\maketitle

\section{Introduction}

The principles of self-organization (SO), implying the autonomous increase of order in a system, is omnipresent in technological and biological entities \cite{nicolis_self_org,cross_pattern}.
 Typically, SO is driven by gradients in thermodynamic potentials so that corresponding processes occur under nonequilibrium conditions. During the organizational process, small fluctuations present in an initially disordered system grow selectively. This results in a long-distance coherent behavior. In pattern-forming systems, this coherence is manifested in characteristic features of
the emerged patterns, which are independent of the exact form of the initial fluctuations. Hence, such characteristics can be used to classify the system.

The analysis of self-organizing systems is also motivated by their frequent occurrence in nature. Among others, SO appears in hydrodynamic instabilities \cite{cross_pattern}, concentration patterns during chemical reactions \cite{epstein_nonlin_chem_dyn}, as well as biological evolution and cellular processes \cite{karsenti_self_org_cell_bio}. Self-organizing systems (SOSs) also find various applications such as in the production of supramolecular structures \cite{lehn_supram_chem}, self-organization of organic semiconductors \cite{choi_self_org_sem}, and producing structured surfaces \cite{wu_so_pattern}.

While there is extensive work addressing the characteristics and dynamics of single SOSs, there is comparably little work on coupled processes. In a similar fashion as the interaction of the components of a single SOS lead to the macroscopic coherent behavior of the SOS itself, the communication between these systems may result in a nontrivial collective behavior as well. So far, to the best of our knowledge, such an analysis has only been performed for SO chemical reactions. Specifically, both theoretical and experimental work has been done for the coupling of chemical oscillators \cite{epstein_nonlin_chem_dyn} and extensively for Turing instabilities. The latter occur in a two-component reaction-diffusion system used as a model system for pattern forming reactions \cite{turing_morpho}. In the work of Yang et al. \cite{yang_id_turing_osci}, two separated liquid films were theoretically analyzed, where the same two-component chemical reaction was taking place in each film. The time dependence of the concentrations was described with the equations proposed by Turing. The coupling between the layers was achieved by allowing one of the reactants to diffuse through the membrane separating the liquids. The numerical analysis of this system predicted the emergence of oscillatory patterns, which do not appear in the single-layer configuration. However, the Turing equations are inappropriate for the complete description of a nonequilibrium system in steady state, since the need for constant supply of reactants and removal of products is not considered \cite{cross_pattern}. Moreover, the emergence of the patterns requires the two reactants to have considerably different diffusion coefficients \cite{epstein_nonlin_chem_dyn}. Despite the difficulties, coupled Turing patterns have been experimentally analyzed, and the emergence of superlattice patterns was observed \cite{berenstein_exp_coupled_turing}. Hence, for the Turing instability it is confirmed that the coupled system has properties significantly different from the individual SOS. This motivates the search for the effects of coupling in other physical systems.

To this end, facilitated by the extensive literature on self-organization in thin liquid films, the coupling of a long-wavelength (or deformational) B\'enard-Marangoni instability with a long-wavelength Rayleigh-Taylor instability is analyzed. Research on instabilities in liquid layers is widespread: next to their importance in coating technologies, they are still rich in unresolved scientific questions, while their theoretical description can nevertheless be performed with the Navier-Stokes equations and other well-established transport equations. Furthermore, both instabilities forming the basis of this article have been examined in detail by numerous authors \cite{oron_lubrication, vanhook_lw_exp_theo, burgess_ceiling_drip}. Herein, the coupling between the layers is achieved by the modulation of heat transfer, while mechanical interaction, e.g., by viscous forces, is negligibly small.  The evolution of coupled liquid instabilities has already been examined \cite{merkt_liquid-liquid, merkt_liquid_liquid-gas, rednikov_two-layered_benard, kumar_faraday, nepomn_multilayer}. However, the purpose of these works was not to illuminate synergistic effects, as the coupling mechanisms were usually quite complicated and a clear separation into individual subsystems was not readily possible. By contrast, for the system under discussion in this article, the dynamics of the films can be described with qualitatively identical evolution equations, facilitating the examination of the effects induced by the coupling. As the evolution equations remain reasonably transparent, one can clearly distinguish the two self-organizing subsystems and immediately isolate the effects which are caused by the coupling alone.

\section{Governing equations}

\subsection{Double-layer configuration}

\begin{figure}
\centerline{\includegraphics[width=\linewidth]{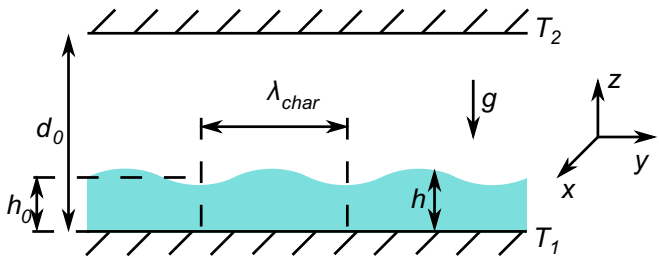}}
\caption{{The double-layer configuration comprising a liquid and a gas film of thicknesses $h_0$ and $d_0-h_0$, respectively, confined between two walls maintained at different but constant temperatures ($T_1>T_2$).}\label{im:one-layer}}
\end{figure}

This section focuses on the evolution equation of the long-wavelength instability for a double-layer system. (Fig. \ref{im:one-layer}) Two distinct types of instabilities exist in liquid films driven by a variation of surface tension with temperature. In corresponding systems, the film with an initial thickness of $h_0$ is typically heated from below, while the free interface of surface tension $\sigma$ is cooled from the top. For thicker films ($h_0 = 0.1-1\text{mm}$), the short-wavelength B\'enard-Marangoni (BM) instability is dominant, for which the characteristic pattern wavelength $\lambda_{char}$ is of the same order as $h_0$, whereas the deformation of the interface is negligible. For the second type, the so-called long-wavelength or deformational BM-instability, $\lambda_{char}$ is much larger than $h_0$, and the interface undergoes significant deformation. Experimentally, this mode of instability is more difficult to observe since the commonly observed short-wavelength mode needs to be suppressed by either a small value of $h_0$ \cite{vanhook_exp} or by using very viscous films such as polymer \cite{troian_polymer-lw} or metal melts \cite{trice_laser-lw}. Equivalently, depending on $h_0$, there exist two types of buoyancy-driven instabilities in liquid films. Of particular interest herein is the long-wavelength Rayleigh-Taylor (RT) instability \cite{vanhook_lw_exp_theo}. As will be described below and except for the sign of the buoyancy term, the evolution equations for the long-wavelength BM instability and the long-wavelength RT instability are qualitatively the same, if in the latter case the film is exposed to a transverse thermal gradient. This was used in Ref. \cite{burgess_ceiling_drip} to prevent the dripping of liquid films from ceilings. 

The derivation of the evolution equation is based on the incompressible Navier-Stokes equations. The detailed analysis is available in multiple papers, therefore we will only summarize the main steps and results. For further details the reader is referred to Refs. \cite{oron_lubrication, vanhook_lw_exp_theo, mathias_nanofilm}. The momentum equations in the bulk of the layer and at the interface read \cite{wang_moving_contact}

\begin{align}
\frac{d\rho \boldsymbol{\mathrm{v}}}{dt}&=-\nabla \cdot \boldsymbol{\mathrm{P}}_{liq} +\rho \boldsymbol{\mathrm{F}}\label{eq:bulk_mom}\\
\frac{d\rho_s \boldsymbol{\mathrm{v}}}{dt}&= - \left(\nabla_s \cdot \boldsymbol{\hat{\mathrm{n}}}\right)\boldsymbol{\hat{\mathrm{n}}}\sigma+\nabla_s\sigma+\left(\boldsymbol{\mathrm{P}}_{liq}-\boldsymbol{\mathrm{P}}_{gas}\right)\boldsymbol{\hat{\mathrm{n}}},\label{eq:surf_mom}
\end{align}

\noindent where $\rho$ is the density, $\boldsymbol{\mathrm{v}}$ is the liquid velocity, and the effects of gravity and other bulk forces are collected in $\boldsymbol{\mathrm{F}}$. The substantial derivative is expressed by $d(.)/dt=\partial(.)/\partial t+{\boldsymbol v}\cdot{\boldsymbol \nabla}(.)$. The stress tensor is denoted by $\boldsymbol{\mathrm{P}}_{liq}$, which is composed of a (scalar) equilibrium pressure $p_{liq}$ and the viscous stress-tensor, namely
\begin{equation}
\begin{aligned} \boldsymbol{\mathrm{P}}_{liq}&=p_{liq}\boldsymbol{\mathrm{I}}+\boldsymbol{\mathrm{E}}^{\nu} &\\
E^{\nu}_{ij}&=-\mu\left(\frac{\partial v_i}{\partial x_j}+\frac{\partial v_j}{\partial x_i}\right) \text{ }i,j\in[1,2,3],
\end{aligned}
\end{equation}

\noindent where $\mu$ is the dynamic viscosity and Newtonian behavior is assumed. The subscripts $1$, $2$ and $3$ represent the $x$, $y$ and $z$ directions, respectively. In Eq.~\eqref{eq:surf_mom}, $\rho_s$ is the surface density, which is usually negligible and will not be considered in this work. The derivative along the surface is denoted by $\boldsymbol \nabla_s$, while $\boldsymbol{\hat{\mathrm{n}}}$ is the normal vector of the surface pointing towards the air layer. Thus the first term on the right hand side of Eq.~\eqref{eq:surf_mom} is the capillary pressure and the second term is the shear stress induced by an inhomogeneity of the surface tension. Compared to liquids, the viscosity of gaseous materials is usually negligible so that the stress-tensor can be approximated by $\boldsymbol{\mathrm{P}}_{gas}=p_{gas}\boldsymbol{\mathrm{I}}$. This assumption is not necessarily valid for very thin layers of gas, where substantial viscous stresses might appear. Nevertheless, the Knudsen number in such thin gas films is no longer small, and additional physical phenomena emerge, such as a velocity slip and a temperature jump at the gas-liquid interface. These effects are beyond the scope of this paper and will not be further discussed.

At the bottom of the liquid layer the no-slip boundary condition is imposed, i.e. $\boldsymbol{\mathrm{v}}|_{z=0}=0$. At the liquid-gas interface the kinematic condition reads $v_z|_{z=h}={\partial h}/{\partial t}+v_x{\partial h}/{\partial x}+v_y{\partial h}/{\partial y}$.

Using the characteristic quantities from Fig. \ref{im:one-layer} we introduce the following nondimensional parameters: $X=x/\lambda_{char}$, $Y=y/\lambda_{char}$, $Z=z/h_0$, and $H=h/h_0$. 
The characteristic flow velocity in the lateral direction $v_c$ is utilized as the scaling factor for the velocities in the form of $V_x=v_x/v_c$, $V_y=v_y/v_c$ and $V_z=(\lambda_{char}/h_0) \cdot v_z/v_c$. The time variable is rescaled according to $\tau=t\cdot v_c/\lambda_{char}$. Finally, $P=p a h_0/\left(\mu v_c\right)$ is the nondimensional form of the pressure, with $a=h_0/\lambda_{char}$. The dimensionless surface tension is denoted by $\Gamma=a\sigma/(\mu v_c)$.

With the nondimensionalized velocities the continuity equation remains qualitatively unchanged
\begin{equation}
\frac{\partial V_x}{\partial X}+\frac{\partial V_y}{\partial Y}+\frac{\partial V_z}{\partial Z}=0.
\end{equation}

The dimensionless bulk momentum equations are given by
\begin{equation}
\begin{aligned}
a \text{Re}\frac{dV_x}{d\tau}=&a^2\left(\frac{\partial^2 V_x}{\partial X^2}+\frac{\partial^2 V_x}{\partial Y^2}\right)+\frac{\partial^2 V_x}{\partial Z^2}-\frac{\partial P}{\partial X}\\
a \text{Re}\frac{dV_y}{d\tau}=&a^2\left(\frac{\partial^2 V_y}{\partial X^2}+\frac{\partial^2 V_y}{\partial Y^2}\right)+\frac{\partial^2 V_y}{\partial Z^2}-\frac{\partial P}{\partial Y}\\
a^3 \text{Re}\frac{dV_z}{d\tau}=&a^4\left(\frac{\partial^2 V_z}{\partial X^2}+\frac{\partial^2 V_z}{\partial Y^2}\right)+a^2\frac{\partial^2 V_z}{\partial Z^2}-\frac{\partial P}{\partial Z}-\frac{\text{Bo}}{\text{Ca}}. \hspace{-20pt}
\end{aligned}
\end{equation}

Furthermore, the dimensionless surface momentum equations read
\begin{equation}
\begin{aligned}
\frac{\partial \Gamma}{\partial X}=&\frac{\partial V_x}{\partial Z}+a^2\frac{\partial H}{\partial X}\frac{\partial V_z}{\partial Z}\\
\frac{\partial \Gamma}{\partial Y}=&\frac{\partial V_y}{\partial Z}+a^2\frac{\partial H}{\partial Y}\frac{\partial V_z}{\partial Z}\\
P=&2a^2\left(\frac{\partial H}{\partial X}\frac{\partial V_x}{\partial Z}+\frac{\partial H}{\partial Y}\frac{\partial V_y}{\partial Z}+\frac{\partial V_z}{\partial Z}\right)-\frac{\nabla_{\|}^2 H}{\text{Ca}}. \hspace{-20pt}
\end{aligned}
\end{equation}

In the long-wavelength limit, $a$ is assumed to be small so that $a^2\ll1$. Furthermore, the characteristic velocity $v_c$ is typically low and the Reynolds number $\text{Re}=\rho h_0 v_c/\mu$ can be assumed to be of order $a$ or smaller. The effect of gravity is captured by the term proportional to $\text{Bo}/\text{Ca}$, where $\text{Bo}=g \rho \lambda_{char}^2/\sigma$ is the Bond number, $\text{Ca}=\mu v_c/(\sigma a^3)$ is the capillary number and $\nabla_{\|}=(\partial/\partial X,\partial/\partial Y)$ is the gradient along the lateral coordinates.

The gradient of the dimensionless surface tension $\nabla_\|\Gamma$ is defined by the variance of the temperature along the interface, i.e. $\nabla_\|\Gamma=$${a}/(\mu v_c)\nabla_\|\sigma=$$-a\sigma_T/(\mu v_c)\nabla_\|T{\big|_{ Z=H}}$. Here, $\sigma$ is assumed to be a linear function of the temperature and $\sigma_T=-{d\sigma}/{dT}$. The nondimensional equivalent of the temperature is defined according to  $\Theta=\left({T-T_2}\right)/\left({T_1-T_2}\right)$, where $T_1$ and $T_2$ are the substrate temperatures of the lower and the upper substrate, respectively, and $T_1>T_2$. Thus the surface tension gradient can be expressed by $\nabla_\|\Gamma=-\text{Ma}\nabla_\|\Theta{\big|_{ Z=H}}$. The dimensionless Marangoni number $\text{Ma}={a\sigma_T(T_1-T_2)}/(\mu v_c)$ characterizes the variance of the surface tension with temperature. This definition of the Marangoni number differs from the one used in the short-wavelength B\'enard-Marangoni instability \cite{vanhook_lw_exp_theo}. In the latter case,  $\text{Ma}^*={\sigma_T\Delta T^*h_0}/({\mu\alpha})$, where $\alpha$ is the thermal diffusivity and $\Delta T^*=T_1-T(h_0)$ is the temperature drop across the liquid layer. 

For the long-wavelength approximation, terms of order $a^2$ and $a^4$ are assumed to be negligible. For $\text{Re} \leq O(a)$, the momentum equations together with the boundary conditions and the continuity equation can be transformed into \cite{mathias_nanofilm}
\vspace{-12pt}

\begin{equation}\label{eq:h_mom_eq}
\begin{aligned}
\frac{\partial H}{\partial \tau}+\nabla_\|\Biggl\{
\frac{H^3}{3\text{Ca}}\Bigl[\nabla_\|\left(\nabla_\|^2H\right)-\text{Bo}\nabla_\|H\Bigr]-&\\
-\text{Ma}\frac{H^2}{2}\nabla_\|\left(\Theta{\big |_{ Z=H}}\right)\Biggr\}&=0
\end{aligned}
\end{equation}

The temperature distribution in the liquid and gas layers can be calculated with the energy equation. In the long-wavelength approximation of the double-layer configuration the P\'eclet numbers $\text{Pe}_{l}$$=$$\text{Re}\cdot \text{Pr}$$=$$h_0 v_c/\alpha$ of the liquid and $\text{Pe}_{g}$$=$$\text{Re}_g\cdot \text{Pr}_g$$=$$h_{0,g} v_{c,g}/\alpha_g$ of the gas layers are assumed to be small, at least to order $a$. The quantities $\alpha$ and $\alpha_g$ stand for the thermal diffusivity of the liquid and gas layers respectively. In nondimensional form, the energy equation reads 
\vspace{-12pt}

\begin{equation}\label{eq:full_energy_eq}
a \text{Pe}_{l/g}\frac{d\Theta}{d\tau}=a^2\left(\frac{\partial^2 \Theta}{\partial X^2}+\frac{\partial^2 \Theta}{\partial Y^2}\right)+\frac{\partial^2 \Theta}{\partial Z^2},
\end{equation}

\noindent where $\text{Pe}_{l/g}$ is either $\text{Pe}_l$ or $\text{Pe}_g$, depending on which phase Eq. \eqref{eq:full_energy_eq} refers to. The first terms on the right-hand side and the left-hand side of Eq.~\eqref{eq:full_energy_eq} can be omitted, as they are proportional to $a^2$. Assuming continuity in the temperature and in the heat flux at the interface, the temperature at $Z=H$ is given by:
\vspace{-12pt}

\begin{equation}
\Theta\big|_{Z=H}=\frac{D_0-H}{D_0+(1/\kappa-1)H},
\end{equation}

\noindent where $\kappa=k_{liq}/k_{gas}$ denotes the ratio of the heat-conductivities of the two layers. The distance between the two substrates is given by $D_0=d_0/h_0$.

Substituting this into Eq.~\eqref{eq:h_mom_eq} leads to a nonlinear partial-differential equation for $H$. For the linear analysis one assumes that the deviation from the equilibrium configuration ($H=1$) is small so that $H=1+\Delta H$, where $\Delta H^2<<1$. With this the equation for $\Delta H$ takes the following form:

\vspace{-12pt}
\begin{equation}\label{eq:one_layer_linearized}
\begin{aligned}
\frac{\partial \Delta H}{\partial \tau}&+\frac{\frac{1}{\kappa} D_0 \text{Ma}}{2\left(D_0+1/\kappa-1\right)^2}\nabla_\|^2\Delta H+\\
&+\frac{1}{3\text{Ca}}\nabla_\|^2\left(\nabla_\|^2\Delta H-\text{Bo}\Delta H\right)=0.
\end{aligned}
\end{equation}

\subsection{Triple-layer configuration}

\begin{figure}
\includegraphics[width=\linewidth]{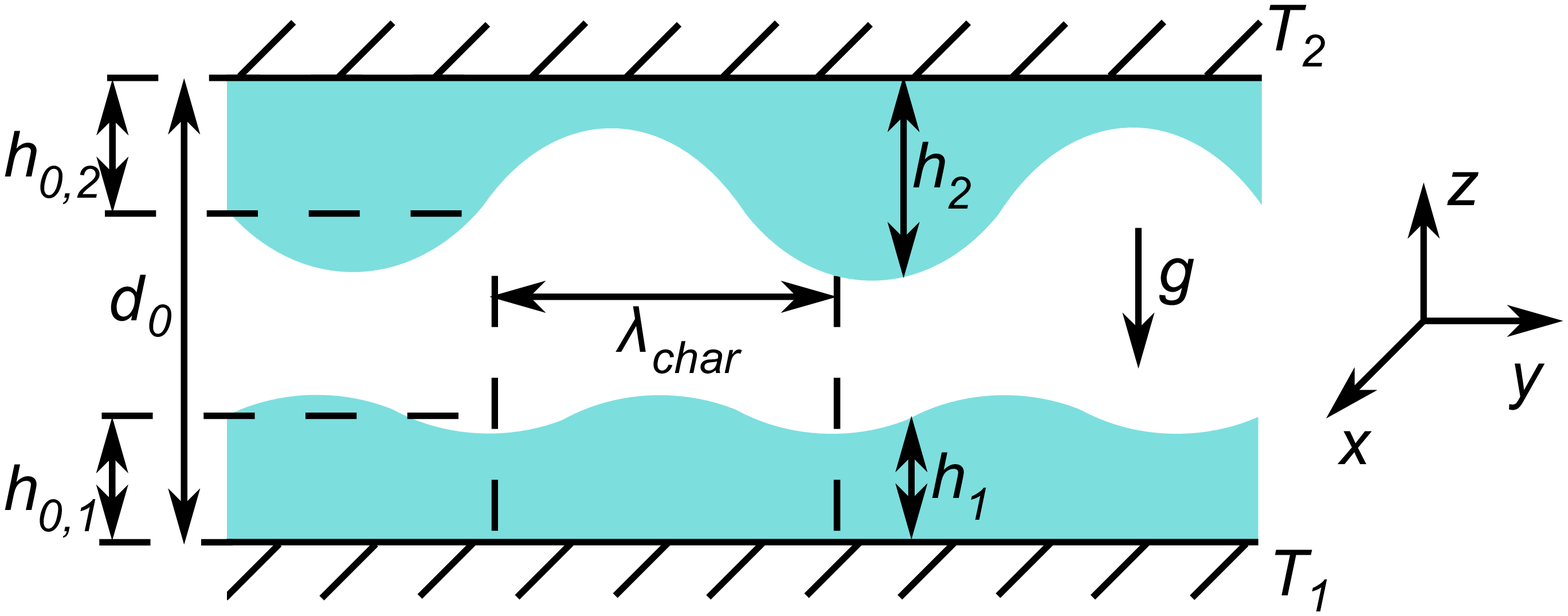}
\caption{Schematic of the triple-layer configuration with initial liquid thicknesses of $h_{0,1}$, $h_{0,2}$ and plate separation $d_0$.}\label{im:double-layer}
\end{figure}

In the system shown in Fig.~\ref{im:one-layer}, thermocapillarity destabilizes the film, while gravity stabilizes it. If the film is placed on the upper substrate, gravity destabilizes it while thermocapillarity stabilizes it. This circumstance was utilized in the thermocapillarity-driven dripping prevention from ceilings \cite{burgess_ceiling_drip}. Note that this system can be described by simply inverting the sign in front of the gravity term in Eq.~\eqref{eq:h_mom_eq}. Inspired by the work of Srivastava et al. \cite{sharm_electric_int}, in the following, two films opposite to each other on separated substrates with a thin gas layer in between are considered (Fig.~\ref{im:double-layer}). As for the two-layer system, the lower substrate is hotter than the upper one so that the lower film is subjected to the gravity-stabilized long-wavelength BM instability, whereas the upper film is subjected to the thermocapillarity-stabilized long-wavelength RT instability. Experimental verification of the core principle of coupled self-organization based on a similar system as the one presented herein is on its way. In that context, it has to be noted that, for the coupled system, the in situ measurement of the individual film height distributions without disturbing the coupling is experimentally very challenging. In the focus of the subsequent analysis are the consequences of thermal coupling between the two films which cannot be observed for the isolated systems.

Due to the negligibly small viscosity of the gas, there is no direct mechanical connection between the layers. Therefore, the momentum equations of the two systems remain independent and the coupling will only appear through the energy equation. As will be shown, the layers are thermally coupled in a nontrivial way, as the deformation of one liquid layer changes the local surface temperature of the other one.

Hereafter it is assumed that the long-wavelength \nolinebreak approximation is valid for both liquids. Consequently  $\left(h_{0,1}/\lambda_{char,1}\right)^2$$\ll$$1$ and $(h_{0,2}/\lambda_{char,2})^2$$\ll$$1$, where $h_{0,i}$ and $\lambda_{char,i}$ represent the initial thicknesses and the characteristic deformation wavelengths of the two liquid layers, respectively. Hence, the evolution equations for the dimensionless liquid thicknesses $H_1=h_1/h_{0,1}$ and $H_2=h_2/h_{0,2}$ can be derived in an equivalent fashion as for Eq.~\eqref{eq:h_mom_eq}. Moreover, we suppose that the thicknesses of the two liquid layers are of the same order of magnitude, and that this assumption remains valid for the characteristic wavelengths too. Therefore, the ratio of the initial thicknesses $\chi=h_{0,2}/h_{0,1}=O(1)$ and the same scaling lengths can be used in the lateral direction. For the calculation of the dimensionless groups, $h_{0,1}$ is used for nondimensionalization. Accordingly, the film evolution equations of both liquid layers are given by
\vspace{-12pt}

\begin{align}
&\begin{aligned}\frac{\partial H_1}{\partial \tau}+\nabla_\|\Biggl\{\frac{H_1^3}{3\text{Ca}_1}\Bigl[\nabla_\|\left(\nabla_\|^2H_1\right)-\text{Bo}_1\nabla_\|H_1\Bigr]-&\\
-{\text{Ma}_1}\frac{H_1^2}{2}\nabla_\|\left(\Theta{\big |_{ Z=H_1}}\right)\Biggr\}&=0
\end{aligned}\label{eq:h_mom_eq_double_1}\\
&\text{and}\nonumber \\
&\begin{aligned}
\frac{\partial H_2}{\partial \tau}+\nabla_\|\Biggl\{
\chi^3\frac{H_2^3}{3\text{Ca}_2}\Big[\nabla_\|\left(\nabla_\|^2H_2\right)
+\text{Bo}_2\nabla_\|H_2\Big]-&\\
-\chi{\text{Ma}_2}\frac{H_2^2}{2}\nabla_\|\left(\Theta{\big |_{ Z=D_0-H_2}}\right)\Biggr\}&=0.
\end{aligned}\label{eq:h_mom_eq_double_2}
\end{align}

\noindent The effect of the different liquid thicknesses are captured by $\chi$ alone. Using the one-dimeansional heat diffusion equation, derived in the previous section, the continuity of the temperature and the heat flux density at the interfaces leads to the following expressions for the interfacial temperatures
\begin{equation}\label{eq:dimles_temp_2}
\begin{aligned}
\Theta\big|_{Z=H_1}& =\frac{-H_1}{\kappa_1(D_0-H_1-\chi H_2)+H_1+\frac{\kappa_1}{\kappa_2}\chi H_2}+1,\\
\text{and}\\
\Theta\big|_{Z=D_0-H_2}& =\frac{\chi H_2}{\kappa_2(D_0-H_1-\chi H_2)+\chi H_2+\frac{\kappa_2}{\kappa_1} H_1},
\end{aligned}
\end{equation}
\noindent where $\kappa_i=k_{liq,i}/k_{gas}$. The surface gradients of the interfacial temperatures are

\vspace{-12pt}
\begin{equation}\label{eq:epsilon_def}
\begin{aligned}&\nabla_\|\Theta|_{H_1}=-\epsilon_1 \nabla_\|H_1-\epsilon_2\nabla_\|H_2 \text{, where} \\
&\displaystyle \epsilon_1=\frac{\kappa_1\Big[D_0-\chi H_2 \left(1-\frac{1}{\kappa_2}\right)\Big]}{\Big[\kappa_1\left(D_0-H_1-\chi H_2\right)+H_1+\frac{\kappa_1}{\kappa_2}\chi H_2\Big]^2},\\
&\displaystyle \epsilon_2=\frac{\kappa_1\chi H_1\left(1-\frac{1}{\kappa_2}\right)}{\Big[\kappa_1\left(D_0-H_1-\chi H_2\right)+H_1+\frac{\kappa_1}{\kappa_2}\chi H_2\Big]^2}
\end{aligned}
\end{equation}

\vspace{-12pt}
\begin{equation}\label{eq:phi_def}
\begin{aligned}
&\text{and } \nabla_\|\Theta|_{D_0-H_2}=\phi_1 \nabla_\|H_1+\phi_2\nabla_\|H_2\text{, where} \\
&\displaystyle \phi_1=\frac{\kappa_2 \chi H_2\left(1-\frac{1}{\kappa_1}\right)}{\Big[\kappa_2\left(D_0-H_1-\chi H_2\right)+\chi H_2+\frac{\kappa_2}{\kappa_1}H_1\Big]^2},\\
&\displaystyle \phi_2=\frac{\kappa_2 \chi \Big[D_0-H_1\left(1-\frac{1}{\kappa_1}\right)\Big]}{\Big[\kappa_2\left(D_0-H_1-\chi H_2\right)+\chi H_2+\frac{\kappa_2}{\kappa_1}H_1\Big]^2}.
\end{aligned}
\end{equation}

\noindent Note that $\epsilon_1,\epsilon_2,\phi_1,\phi_2>0$ and $\epsilon_1\phi_2>\epsilon_2\phi_1$.

For the linear analysis we introduce the notation $H_1=1+\Delta H_1$ and $H_2=1+\Delta H_2$ and assume again that $\Delta H_1^2,\quad \Delta H_2^2 <<1$. Neglecting the second and higher order terms in $\Delta H_1$ and $\Delta H_2$ leads to the linearized evolution equations, reading

\begin{equation}\label{eq:non_dim_grav}
\begin{aligned}
\frac{\partial \Delta H_1}{\partial \tau}+\left( \frac{\epsilon_1 \text{Ma}_1}{2}-\frac{\text{Bo}_1}{3\text{Ca}_1}\right)\nabla_\|^2 \Delta H_1+&\\
+\frac{\epsilon_2 \text{Ma}_1}{2}\nabla_\|^2 \Delta H_2+\frac{1}{3\text{Ca}_1}\nabla_\|^4 \Delta H_1&=0, \\
\frac{\partial \Delta H_2}{\partial \tau}+\left( -\frac{\phi_2\chi \text{Ma}_2}{2}+\frac{\chi^3 \text{Bo}_2}{3\text{Ca}_2}\right)\nabla_\|^2 {\Delta H_2}-&\\
- \frac{\phi_1 \chi \text{Ma}_2}{2}\nabla_\|^2 {\Delta H_1}+\frac{\chi^3}{3\text{Ca}_2}\nabla_\|^4 {\Delta H_2}&=0.
\end{aligned}
\end{equation}

\noindent Here, $\epsilon_1$, $\epsilon_2$, $\phi_1$ and $\phi_2$ are evaluated at $H_1 = 1$ and $H_2 = 1$, respectively.

\section{Linear and numerical analysis of the evolution equations}

In this section, the properties of Eq.~\eqref{eq:non_dim_grav} are analyzed. This will provide useful information for the qualitative and quantitative description of the emerging patterns as well as the stability behavior of the SOS \cite{cross_pattern}.

Equation \eqref{eq:non_dim_grav} indicates that the linear evolution of the liquid layers is qualitatively similar. The main difference lies in the interchanged roles of the thermocapillary and gravitational force. While the $Ma_1$-term in the first equation destabilizes the lower layer by giving positive feedback to a deformation, the $Ma_2$-term in the second layer stabilizes the upper layer by damping deformations. The opposite is true for the gravitational body force, i.e., it stabilizes the lower layer but destabilizes the upper layer. 

For a better understanding of some of the results presented later in this work, it is helpful to consider a simplified version of the triple-layer configuration, where the upper liquid layer is assumed to be rigid. (${\Delta H_2}=0$). In this case the second equation in Eqs.~\eqref{eq:non_dim_grav} turns into an identity, while the first one simplifies to

\vspace{-12pt}
\begin{equation}\label{eq:upper_rigid}
\displaystyle \frac{\partial H_1}{\partial \tau}+\frac{\frac{1}{\kappa_1} D_0 \text{Ma}_{eff}}{2\left(D_0+\frac{1}{\kappa_1}-1\right)^2}\nabla_{\|}^2H_1+\frac{1}{3\text{Ca}_1}\nabla_{\|}^4H_1=0,
\end{equation}

with the effective Marangoni number

\vspace{-12pt}
\begingroup
\fontsize{10pt}{8pt}\selectfont
\begin{equation}\label{eq:maeff}
\text{Ma}_{eff}=\left({\frac{\epsilon_1\text{Ma}_1}{2}-\frac{\text{Bo}_1}{3\text{Ca}_1}}\right)\left[{\frac{\frac{1}{\kappa_1} D_0}{2\left(D_0+(\frac{1}{\kappa_1}-1)\right)^2}}\right]^{-1}.
\end{equation}
\endgroup

Equation \eqref{eq:upper_rigid} has the form of a linearized evolution equation of a system with only one liquid layer in the absence of gravity, where $Ma_{eff}$ serves as the Marangoni number. Thus, it is an equivalent of Eq.~\eqref{eq:one_layer_linearized} and the results already available for this type of instability \cite{oron_lubrication} are directly applicable. In particular, the system will always be linearly unstable if $\text{Ma}_{eff}>0$, which is equivalent to $\epsilon_1\text{Ma}_1/2-{\text{Bo}_1}/(3\text{Ca}_1)>0$. In the limit of  $\chi =0$ this criterion is identical to the results obtained by VanHook et al. \cite{vanhook_lw_exp_theo}. The same train of thought can be applied to a system where the lower layer is rigid. In this case the upper one is unstable if $-\phi_2 \text{Ma}_2/2+\chi^2\text{Bo}_2/(3 \text{Ca}_2)>0$.

In general, for both layers being mobile, Eq.~\eqref{eq:non_dim_grav} can be solved using the Fourier transforms of the deformation variables
\vspace{-12pt}

\begin{equation}\label{eq:fourier_form}
\begin{aligned} {\Delta H_n}&(\tau,X,Y)=\\
=&\frac{1}{2\pi}\int_{-\infty}^{\infty}{\widetilde{\Delta H}_n(\tau,q_{x},q_{y})e^{i(q_{x}X+q_{y}Y)}dq_{x}dq_{y}},
\end{aligned}
\end{equation}

\noindent where $n\in[1;2]$ indicates the two liquid layers. The dimensionless wavenumbers (scaled by $a/h_0$) in the $x$ and $y$ direction are denoted by $q_x$ and $q_y$. The back substitution to the linearized evolution equations gives an equation for every Fourier component. As Eq.~\eqref{eq:non_dim_grav} is linear, the different Fourier modes will be independent of each other. Introducing $q=\left(q_x^2+q_y^2\right)^{1/2}$, the transformed equations read
\vspace{-12pt}

\begin{equation}\label{eq:non_dim_matrix_grav}
\frac{\partial}{\partial\tau}\begin{pmatrix} {\widetilde{\Delta H}_1}\\ {\widetilde{\Delta H}_2}\end{pmatrix}=\boldsymbol{\mathrm{M}}
\begin{pmatrix} {\widetilde{\Delta H}_1}\\ {\widetilde{\Delta H}_2}\end{pmatrix},
\end{equation}

\vspace{-12pt}
\begin{align*}
&M_{1,1}=\left( \frac{\epsilon_1 \text{Ma}_1}{2}-\frac{\text{Bo}_1}{3\text{Ca}_1}\right)q^2-\frac{1}{3\text{Ca}_1}q^4,\\
&M_{1,2}=\frac{\epsilon_2\text{Ma}_1}{2}q^2, \quad \quad M_{2,1}=
-\frac{\phi_1\chi \text{Ma}_2}{2}q^2, \\
&M_{2,2}=\left( -\frac{\phi_2\chi \text{Ma}_2}{2}+\frac{\chi^3\text{Bo}_2}{3\text{Ca}_2}\right)q^2-\frac{\chi^3}{3\text{Ca}_2}q^4.
\end{align*}

\noindent The general solution of this system of linear differential equations is 
\begin{equation}\label{eq:general_fourier_sol}
\begin{aligned}
\displaystyle &\begin{pmatrix}{\widetilde{\Delta H}_1}(\tau,q_x,q_y)\\ {\widetilde{\Delta H}_2}(\tau,q_x,q_y)\end{pmatrix}=\\
&=\begin{pmatrix}{\widetilde{\Delta H}_{1+}}(q)\\ {\widetilde{\Delta H}_{2+}}(q)\end{pmatrix}e^{\omega_+\tau}+\begin{pmatrix}{\widetilde{\Delta H}_{1-}}(q)\\ {\widetilde{\Delta H}_{2-}}(q)\end{pmatrix}e^{\omega_-\tau}.
\end{aligned}
\end{equation}

In Eq.~\eqref{eq:general_fourier_sol}, $\omega_+(q)$ and $\omega_-(q)$ are the eigenvalues of the matrix $\boldsymbol{\mathrm{M}}$ in Eq.~\eqref{eq:non_dim_matrix_grav}, while $\left(\widetilde{\Delta H}_{1+}(q), \widetilde{\Delta H}_{2+}(q)\right)^T$ and $\left(\widetilde{\Delta H}_{1-}(q), {\widetilde{\Delta H}_{2-}}(q)\right)^T$ are the respective eigenvectors.

\subsection{Linearized equations for identical layers}

The exact formulas for the eigenvectors and eigenvalues are involved and are functions of many independent parameters. One arrives at considerably simpler formulas if the layers are identical. ($\epsilon_1=\phi_2$, $\epsilon_2=\phi_1$, $\text{Ca}_1=\text{Ca}_2\equiv \text{Ca}$, $\text{Ma}_1=\text{Ma}_2\equiv \text{Ma}$, $\text{Bo}_1=\text{Bo}_2\equiv \text{Bo}$ and $\chi=1$.). For simplification, the dimensionless time is rescaled according to $\hat{\tau}=\tau / 3\text{Ca}$. The resulting eigenvalues are

\vspace{-12pt}
\begin{equation}\label{eq:ident_eigval}
\displaystyle\omega_{\pm}=-q^4 \pm q^2 \cdot 3\text{Ca}\sqrt{\left( \frac{\epsilon_1 \text{Ma}}{2}-\frac{\text{Bo}}{3\text{Ca}}\right)^2-\left(\frac{\epsilon_2 \text{Ma}}{2}\right)^2}
\end{equation}

\noindent For this type of system the real part of the second eigenvalue $\text{Re}\left(\omega_-\right)$ is always negative. Therefore the corresponding modes are damped and will not have any effect on the long-term evolution of the system. To analyze the other eigenvalue from Eq.~\eqref{eq:ident_eigval} we define
\begin{equation}
s=\left(\frac{3}{2}\epsilon_1\text{Ca} \text{Ma}-\text{Bo}\right)^2-\left(\frac{3}{2}\epsilon_2\text{Ca} \text{Ma}\right)^2.
\end{equation}

\begin{figure}
\centerline{\includegraphics[width=\linewidth]{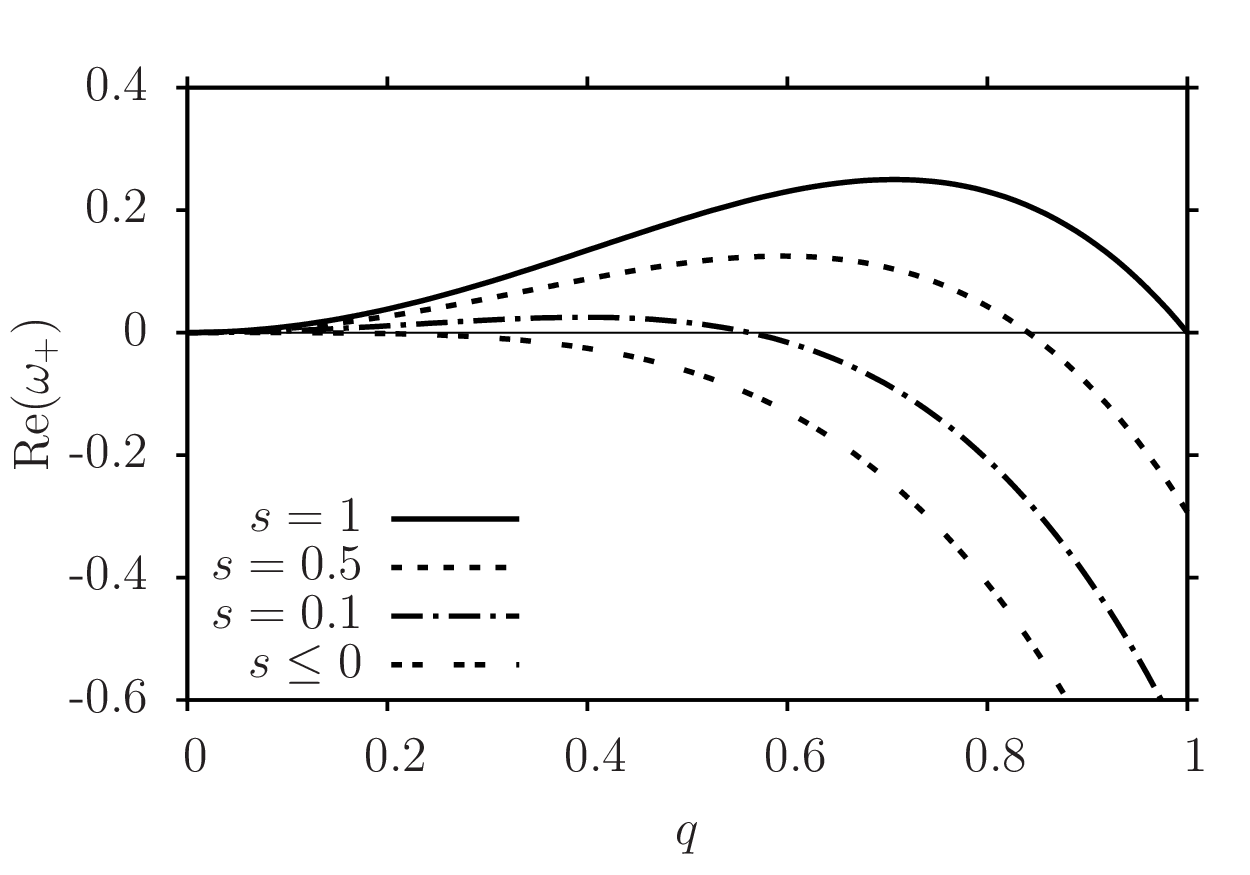}}
\caption{{The growth rate function of the linear stability analysis as a function of the dimensionless wavenumber.}\label{im:growthrate}}
\end{figure}

\noindent Further examination of Eq.~\eqref{eq:ident_eigval} indicates that for $s>0$ the system is unstable to small perturbations, as there exists a range of wavenumbers $[q_{min},q_{max}]$ where $\text{Re}(\omega_+)\geq0$. Moreover, the deformations of the film are not oscillatory since $\omega_+$ is a real number in this region. By setting the left-hand side of Eq.~\eqref{eq:ident_eigval} to zero one obtains the marginally stable wavenumbers, $q_{max}=\sqrt[4]{s}$ and $q_{min}=0$. The marginal stability of the $q=0$ mode can also be understood intuitively: a growth rate different from zero would imply a uniform thickening or thinning of the film, changing the volume of the layer, which violates mass conservation \cite{cross_pattern}. In Fig.~\ref{im:growthrate} the growth rate $\text{Re}\left(\omega_+(s)\right)$ as a function of the wavenumber is shown. The aforementioned properties of the pattern formation identify it as a type-II-s instability \cite{cross_pattern}.

In the framework of the linear analysis, the characteristic wavenumber of the emerging pattern is predicted by finding the quantity $q_{char}$ that maximizes the real part of the growth rate. For the identical-layer setup, one finds that, similarly as for the double-layer configuration, $\displaystyle q_{char}=q_{max}/\sqrt{2}$. This defines a larger characteristic wavelength than for the uncoupled system. The corresponding growth rate is $\omega_{char}=s/4$.

For validation of the results, a series of direct numerical simulations of Eq.~\eqref{eq:h_mom_eq_double_1} and Eq.~\eqref{eq:h_mom_eq_double_2} were performed and were compared with the findings of the linear analysis. Different values of $s$ were examined. For the liquid films the properties of a 10cSt silicone oil were used, separated by a layer of air. The material properties are summarized in Table \ref{tab:mat_prop}. The thermal conductivity and thermal diffusivity of the air layer at $50\,^{\circ}\mathrm{C}$ and at atmospheric pressure were approximated by $k_g=27.88\cdot 10^{-3}\text{W}\text{m}^{-1}\text{K}^{-1}$ and $\alpha_{g}=2.57\cdot10^{-5}\text{m}^2/\text{s}$ \cite{vdi}. For this analysis the thicknesses of both the air and the liquid layers were set to $h_{0,1}=100\mu \text{m}$. Systems with slightly different film heights were also analyzed in a second series of simulations. This will be addressed later. The value of $s$ was controlled by the temperature difference between the two substrates. By varying the latter between $7.5\text{K}$ and $20.5\text{K}$, $s$ varies between $0.1$ and $3.5$. Within this region, the theoretical expectations of the characteristic wavelengths are in the range $[64\cdot h_{0,1}, 158\cdot h_{0,1}]$. It follows that $a^2\ll1$. Furthermore, in the third column of Table {\ref{tab:dim_quant}}, the typical values of the dimensionless groups characterizing the liquid layers are summarized. In order to predict their values, $q_{char}$ was calculated based on the linear stability analysis and was transformed back to its dimensional form according to the lateral scaling length used in the simulations. The calculation of $\text{Re}$, $\text{Ma}$, $\text{Ca}$ and $\text{Pe}$ requires the characteristic velocity $v_c$. Thus they could be only computed in the unstable ($s>0$) region. The characteristic velocity was approximated with the results of the linear analysis, and the method used to estimate it is discussed in the appendix. As apparent from the table, the assumptions underlying the lubrication approximation are valid for the simulated systems. Furthermore, the conventional Marangoni number describing the onset of the short-wavelength B\'enard-Marangoni is below the critical value of 80 \cite{pearson_sw}. Hence, this instability is not present in our system and all unstable modes have a long-wave character.

For the numerical simulations the finite-element method was applied with quadratic Lagrangian shape functions. They were conducted with Comsol 4.3 \cite{comsol} using the Matlab Livelink environment. The lateral scaling length was set to $10h_{0,1}$. As the physical behavior of the system is independent of the scaling parameters this does not have any effect on the evolution of the system. The simulated domain was a square with side lengths of $15\left(2 \pi\right)/q_{char}$. As $q_{char}$ is dependent on $s$, the simulation area was not the same for the different temperatures. Each square was divided into $120\times120$ cells. Hence, according to the Nyquist-sampling theorem, the maximal numerically resolvable wavenumber is $4q_{char}$ if the domain was discretized with a finite-difference scheme. Finite-element discretization with high order interpolation functions exceed this resolution. In any case, this resolution is more than sufficient because, according to the linear analysis, the largest unstable wavenumber is expected to be $q_{max}=\sqrt{2}q_{char}$ and higher wavenumbers will probably be damped. Nevertheless, $q_{max}$ is only an approximation, and the fine meshing should allow for possible deviations to appear in the simulations. Periodic boundary conditions were applied at the edges of the simulation domain so that, as a consequence of the finite size of the rectangle, the numerically available wavenumbers in the $x$- and $y$-direction are multiples of $q_{char}/15$. All simulations started from a near-equilibrium state with $H_1=1+\xi_1(x,y)$ and $H_2=1+\xi_2(x,y)$, where $\xi_1$, $\xi_2$ are white noise perturbations with amplitudes of $5\cdot10^{-2}$. The simulations covered a time span of $\Delta\hat{\tau}=2.5/\omega_{char}$. The maximal time step was set to $0.005/\omega_{char}$. At the final time step the relative deformation of the layers was found to be between $15\%$ and $30\%$. 

All calculations were performed on a Dell Precision T7500 workstation running Cent OS 5.8. Grid independency studies were conducted by varying the mesh size between $30\times 30$ and $180\times 180$. It was found that, beyond a grid resolution of $105 \times 105$ the numerical results became practically indistinguishable from each other. 

\begin{table}
  \centering
  \begin{ruledtabular}
  \begin{tabular}{c  M  c}
    Parameter & Notation & Value\\
    	\thickhline
    	\rule{0pt}{2.5ex}	Density & $\rho$ &$940 \frac{\text{kg}}{\text{m}^3}$\\[1.25ex]
	Surface tension & $\sigma$ &$19 \cdot10^{-3} \frac{\text{N}}{\text{m}}$\\[1.25ex]
	Surface tension coefficient & $\sigma_T$ &$6.9\cdot 10^{-5}\frac{\text{N}}{\text{mK}}$\\[1.25ex]
	Kinematic viscosity & $\nu$ &$1.02 \cdot 10^{-5} \frac{\text{m}^2}{\text{s}}$\\[1.25ex]
	Thermal conductivity & $k$ &$13.3 \cdot 10^{-2} \frac{\text{W}}{\text{mK}}$\\[1.25ex]
	Thermal diffusivity & $\alpha$ & $10^{-7}\frac{\text{m}^2}{\text{s}}$
    \end{tabular}%
    \end{ruledtabular}
    \caption{ Material properties of the considered silicone oil at $50 \,^{\circ}\mathrm{C}$ \cite{vanhook_lw_exp_theo}.}
  \label{tab:mat_prop}
\end{table}

\begin{table}
\centering
\begin{ruledtabular}
  \begin{tabular}{L M M M}
    & Formula & First series & Second series\\
	\thickhline
	\rule{0pt}{4ex}Reynolds & $\displaystyle\frac{\rho h_0 v_c}{\mu}$ & $6.21\cdot10^{-4}$- -$5.33\cdot10^{-3}$ & $1.53\cdot10^{-4}$- -$5.17\cdot10^{-3}$ \\
	\rule{0pt}{3.5ex}Marangoni & $\displaystyle\frac{a\sigma_T(T_2-T_1)}{\mu v_c}$ & $4.17$ - $5.45$ & $4.45$ -$10.1$\\
	\rule{0pt}{3ex}$Ma^*$ & $\displaystyle\frac{\sigma_T \Delta T^* h_{0,1}}{\mu \alpha}$ & $19.2$ - $52.0$ & $9.9$ - $59.6$ \\	
	\rule{0pt}{3ex}Capillary & $\displaystyle\frac{\mu v_c}{\sigma a^3}$ & $75.2$ - $126.0$ & $36.4$ - $110.8$\\
	\rule{0pt}{3.7ex}Galileo & $\displaystyle\frac{\rho g h_{0,1}^3}{\mu \alpha}$ & $9.62$ & $16.62$\\
	\rule{0pt}{3.7ex}Bond & $\displaystyle\frac{g \rho \lambda_{char}^2}{\sigma}$ & $20.49$ - $121.2$ & $36.3$ - $303.4$\\
	\rule{0pt}{3.3ex}Prandtl & $\displaystyle\frac{\mu}{\rho \alpha}$ & 102 & 102\\
	\rule{0pt}{3.3ex}P\'eclet (oil) & $\displaystyle\frac{h_0 v_c}{\alpha}$ & $6.33\cdot10^{-2}$ - $5.43\cdot10^{-1}$ & $1.56\cdot10^{-2}$ - $5.27\cdot10^{-1}$\\
	\rule{0pt}{3.3ex}P\'eclet (air) & $\displaystyle\frac{h_0 v_c}{\alpha_{g}}$ & $2.47\cdot10^{-4}$ - $2.12\cdot10^{-3}$ & $6.08\cdot10^{-5}$ - $2.05\cdot10^{-3}$
	 \end{tabular}%
	 \end{ruledtabular}
   \caption{ The dimensionless groups characterizing the governing equations and their values for the two series of numerical simulations. In the first series a parametric sweep in $s$ was performed, while in the second one the parametric sweep was conducted for $r$.}
  \label{tab:dim_quant}
\end{table}

\begin{figure}
\includegraphics[width=\linewidth]{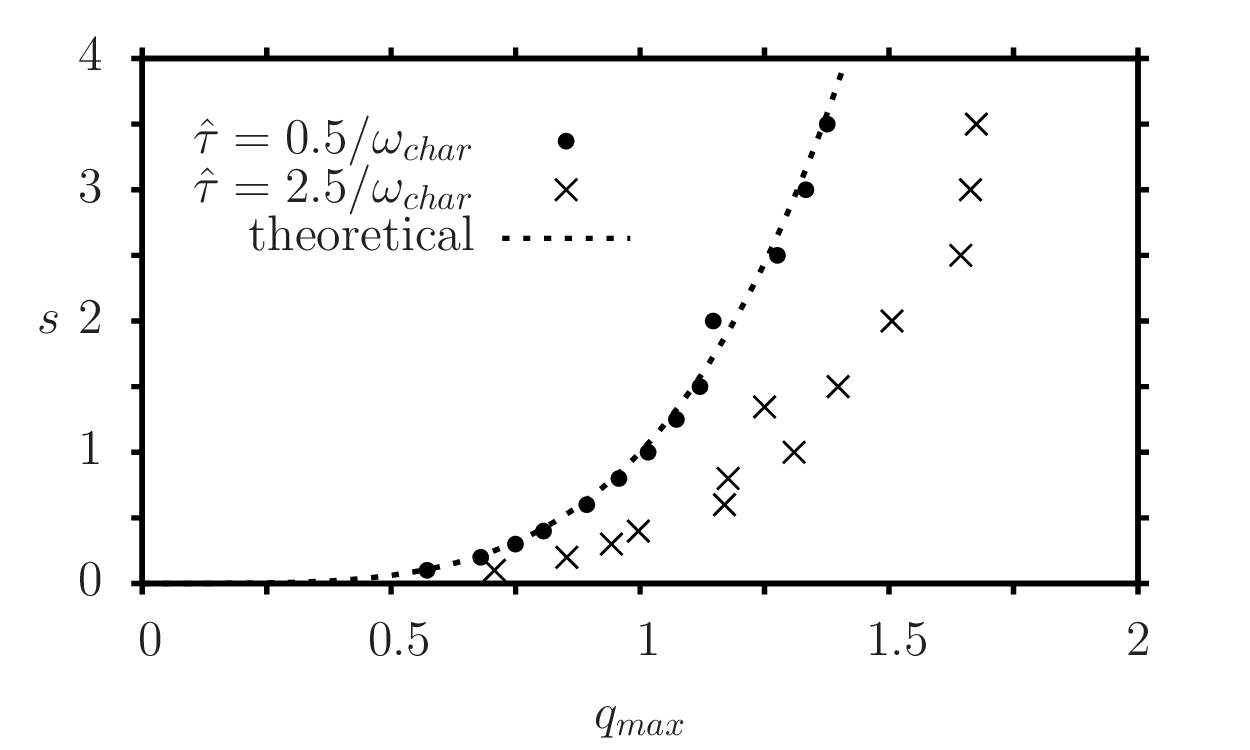}
\caption{The theoretical neutral stability curve and the corresponding numerical results at $\hat{\tau} =0.5/\omega_{char}$ and $\hat{\tau} =2.5/\omega_{char}$.}\label{im:neutral_stab}
\centerline{\includegraphics[width=\linewidth]{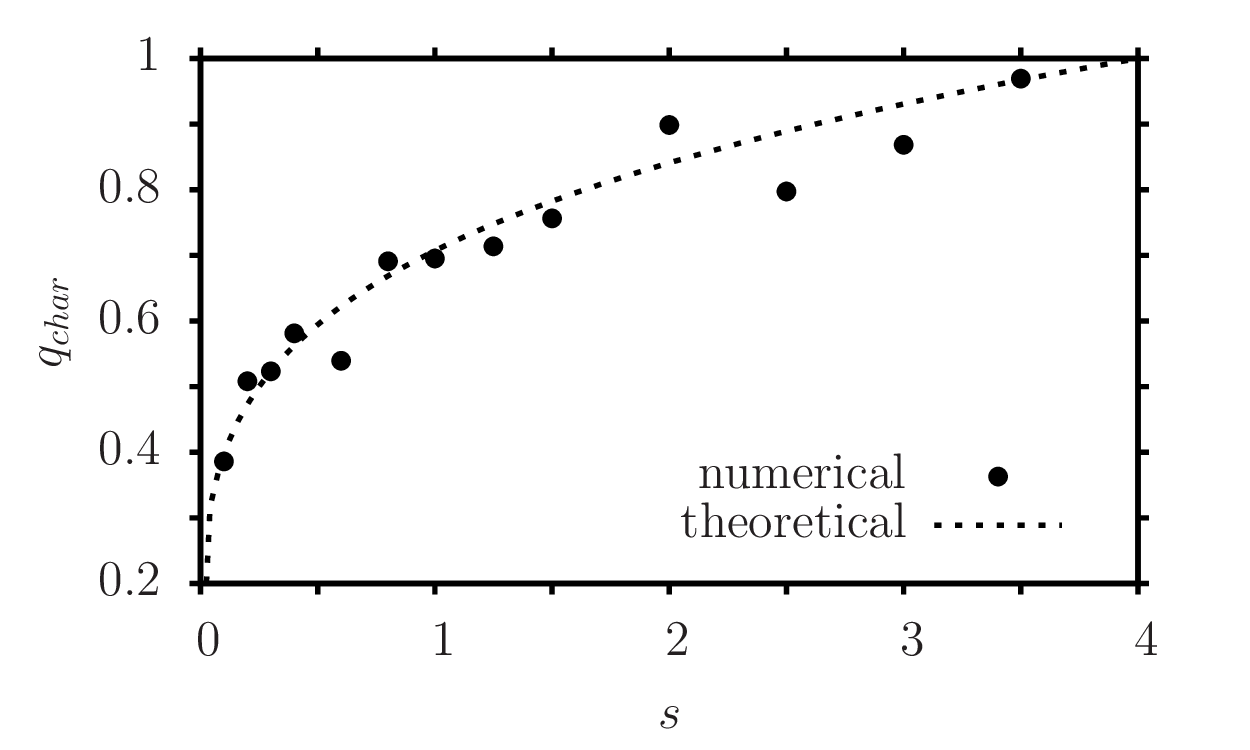}}
\caption{{The theoretical and numerical results ($\hat{\tau} =2.5/\omega_{char}$) for the characteristic wavenumber of the film deformations.}\label{im:char_wave}}
\end{figure}

A comparison between the theoretical predictions and the numerical results for the neutral stability curve $q_{max}(s)$ is given in Fig.~\ref{im:neutral_stab}. The numerical curves were calculated at $\hat{\tau}_1=0.5/\omega_{char}$ and $\hat{\tau}_2=2.5/\omega_{char}$. The data were obtained by analyzing the time evolution of the two-dimensional Fourier transform of the film thickness. A Fourier component was considered to be unstable, if its amplitude increased for ten successive time-steps. The marginally stable wavenumber was approximated by calculating the mean value of two averages. The first one is the directional average of the smallest wavenumbers for which the Fourier component is stable, whereas the second one is the directional average of the largest wavenumbers for which the Fourier component is unstable. At $\hat{\tau}_1=0.5/{\omega_{char}}$ the theoretical expectations and numerical results agree well. The linear method predicts slightly smaller wavenumbers than the numerical results, the largest relative difference between the two is $3.5\%$. By contrast, at $\hat{\tau}_2=2.5/{\omega_{char}}$, the difference between the two solution approaches becomes significant. According to the numerical results $q_{max}$ increases significantly with increasing time. This tendency is present in the whole $[0.5/{\omega_{char}}, 2.5/{\omega_{char}}]$ time interval. This is not expected to be a consequence of numerical inaccuracy, as the spatial resolution is considerably higher than the increased values of $q_{max}$. This behavior suggests that the system becomes more unstable during its time evolution. Further simulations (not
shown for brevity) indicate that the monotonic expansion of the unstable wavenumber region is also present
in double-layer configurations and is thus not a consequence of the coupling. The work of Boos and Thess \cite{boos_cascade} supports this argument, as their numerical analysis of drained regions in a double-layer configuration indicated the destabilization of higher wavenumbers.

At $\hat{\tau}=2.5/\omega_{char}$ the Fourier transforms of $H_1(X,Y)$ and $H_2(X,Y)$ were also used to estimate the characteristic wavenumber of the pattern. This can be approximated by the wavenumber corresponding to the Fourier component with the largest amplitude, which is the one with the largest growth rate within the time range considered. This value was calculated for the patterns of the lower and upper liquid film. The results of the two layers coincided for every simulation except at $s=0.8$, where there was a $1\%$ relative difference between them. As shown in Fig.~\ref{im:char_wave}, the numerical results fit the theoretical expectations well.

From Eqs.~\eqref{eq:epsilon_def}~and~\eqref{eq:phi_def} one can deduce that for identical layers $\epsilon_1 > \epsilon_2$. However, the presence of gravity ($Bo>0$) allows the parameter $s$ to assume negative values. In this case the system is stable, as $\text{Re}(\omega_+)=\text{Re}(\omega_-)=-q^4<0$ for every wavenumber. In the stable regime it is assumed that momentarily appearing film deformations are long-waved in character so that the evolution equations presented before remain valid. For these deformations, $\text{Im}\left(\omega_{\pm}\right)=\pm q^2 \sqrt{-s}$ will give rise to waves appearing on the surface with a dimensionless phase velocity of $V_p=q\sqrt{-s}$ and a group velocity of $V_g=2q\sqrt{-s}$. 

\begin{figure}
\centerline{\includegraphics[width=\linewidth]{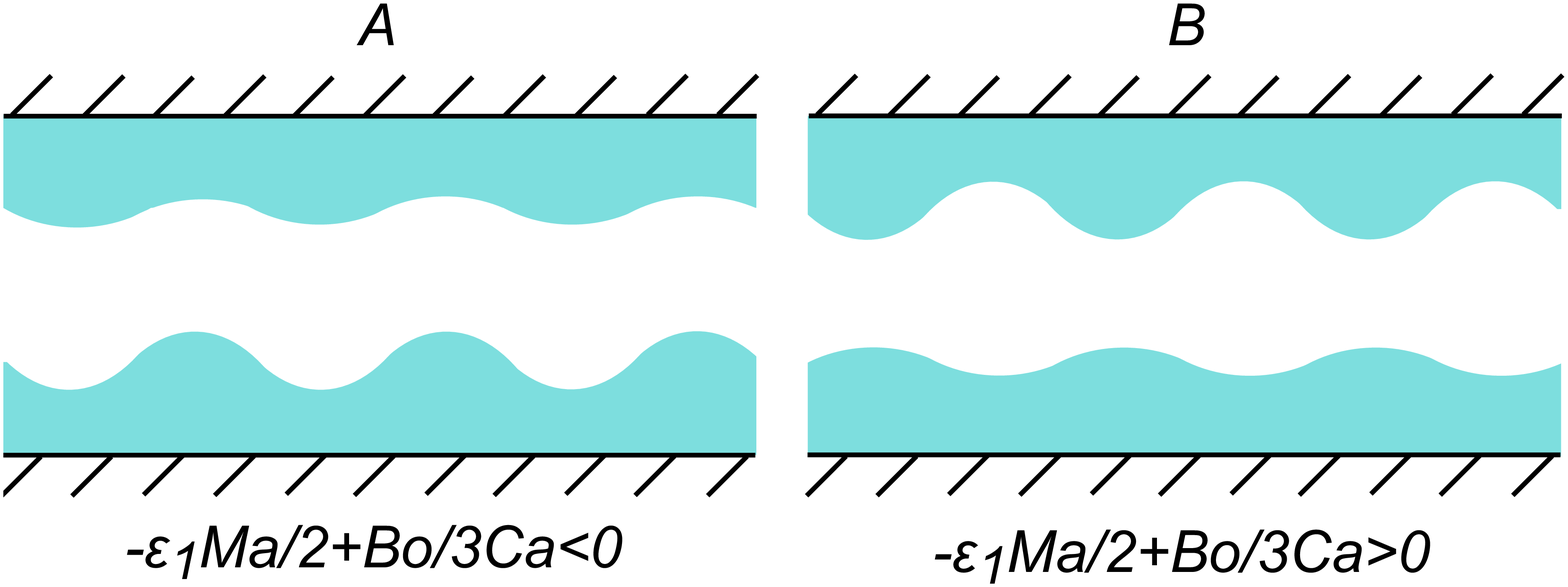}}
\caption{{The triple-layer configurations with a weak effect of gravity (antiphase deformation, $A$) and a strong gravity effect (in-phase deformation, $B$).}\label{im:double_strong_grav}}

\centerline{\includegraphics[width=\linewidth]{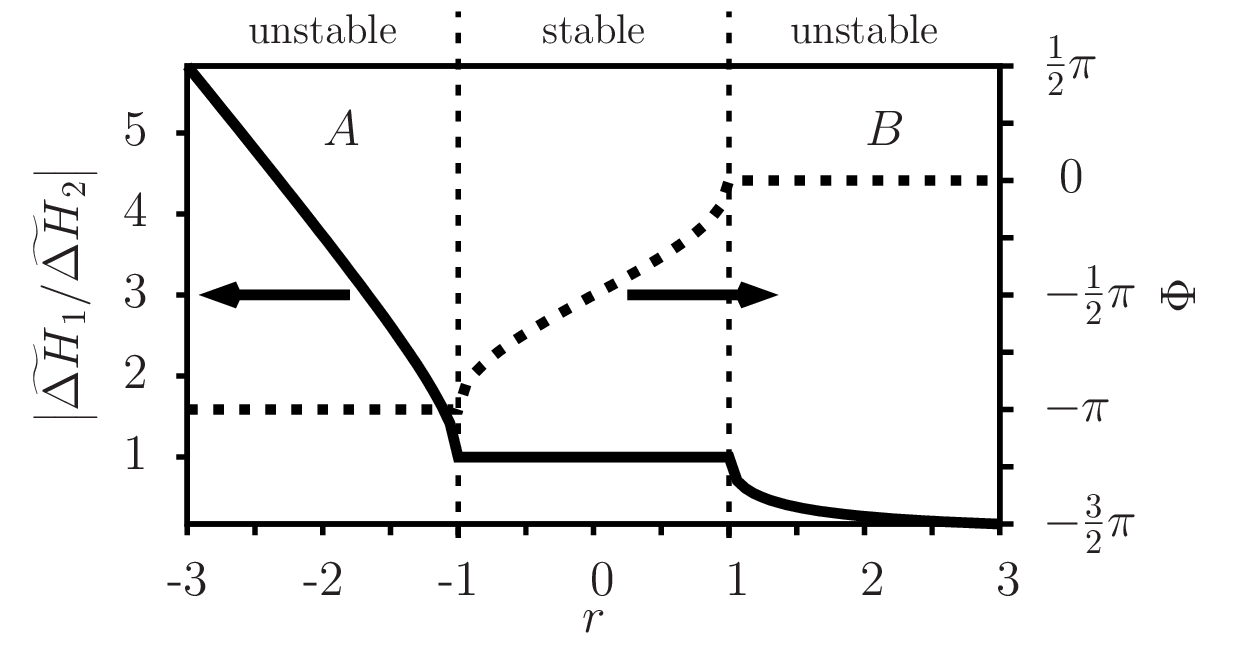}}
\caption{{The ratio of the amplitudes of the two liquid films (left side, solid line) and the phase shift between the patterns (right side, dotted line).}\label{im:summary}}
\end{figure}

These waves only appear because of the coupling of the two layers. From Eq.~\eqref{eq:non_dim_matrix_grav} one can deduce that if the coupling between the layers is set to zero ($\epsilon_2=\phi_1=0$), then $M(1,2)=M(2,1)=0$. In consequence, the eigenvalues of $\boldsymbol{\mathrm{M}}$ are real numbers in the whole parameter range, making it impossible for oscillations to occur in a single liquid film.

For identical layers the eigenvectors of the linearized system described by Eq.~\eqref{eq:non_dim_matrix_grav} are
\vspace{-12pt}

\begin{equation}\label{eq:ident_eigvec}
\begin{aligned}
&\begin{pmatrix}\widetilde{\Delta H}_{1\pm}(q)\\ \widetilde{\Delta H}_{2\pm}(q)\end{pmatrix}= \\
& =\begin{pmatrix}{\displaystyle -\frac{\left(\frac{\epsilon_1 \text{Ma}}{2}-\frac{\text{Bo}}{3\text{Ca}}\right)\pm\sqrt{\left(\frac{\epsilon_1 \text{Ma}}{2}-\frac{\text{Bo}}{3\text{Ca}}\right)^2-\left(\frac{\epsilon_2 \text{Ma}}{2}\right)^2}}{\frac{\epsilon_2 \text{Ma}}{2}}}\\
\\ 1\end{pmatrix},
\end{aligned}
\end{equation}

\noindent that is, they are in fact independent of $q$. As mentioned earlier, $\omega_{-}<0$ for $s>0$. Thus, the effect of the corresponding eigenvector will disappear after a sufficiently long time. Furthermore, since the eigenvectors are the same for every wavenumber, after a sufficiently long time, the ratio of the deformation of the two layers should be $\widetilde{\Delta H}_{1+}/\widetilde{\Delta H}_{2+}$. In other words, the two layers will have exactly the same pattern, with only the magnitude of the deformation differing by a scaling factor. Nevertheless, for the case of $s>0$, there are two qualitatively different configurations: if ${\epsilon_1 \text{Ma}}/{2}-{\text{Bo}}/({3\text{Ca}})>0$, then $\widetilde{\Delta H}_{1+}/\widetilde{\Delta H}_{2+}<0$, i.e., the deformations will be in antiphase. By contrast if ${\epsilon_1 \text{Ma}}/{2}-{\text{Bo}}/({3\text{Ca}})<0$, then $\widetilde{\Delta H}_{1+}/\widetilde{\Delta H}_{2+}>0$; hence the deformations of the two layers are in phase. This is schematically shown in Fig.~\ref{im:double_strong_grav}. 

An intuitive explanation of this behavior can be attained by identifying the effects driving the film evolution. As mentioned earlier, if the upper layer is rigid, then the lower one is unstable if ${\epsilon_1 \text{Ma}}/{2}-\text{Bo}/(3\text{Ca})>0$.  Similarly, if instead of the lower layer the upper layer is the only deformable layer, then it will be unstable if $\epsilon_1 \text{Ma}/{2}-\text{Bo}/(3\text{Ca})<0$. These are two mutually exclusive conditions. Returning to the coupled system, this suggests that, if the former condition is fulfilled, the evolution of the system should be mainly driven by the lower layer. On top of the regions where the lower layer thickens, the surface of the upper layer will heat up. Subsequently the Marangoni flow arising in the upper layer will point away from this location, leading to a locally decreasing thickness of the upper film. This behavior leads to the antiphase configuration depicted at the left-hand side of Fig.~\ref{im:double_strong_grav}. Based on similar arguments, the in-phase evolution can be explained by considering that, in this case, the instability will be mainly driven by the upper layer, i.e., by gravity.

For $s<0$ according to Eq.~\eqref{eq:ident_eigvec}, the magnitudes of the deformations are equal, but with a phase shift $\Phi$ between the two layers.  Formally $\widetilde{\Delta H}_{1+}/\widetilde{\Delta H}_{2+} \equiv e^{i \Phi}$, where $\Phi$ is independent of the wavenumber.

The phase shift behavior and amplitude ratios can be described in terms of the parameter

\vspace{-12pt}
\begin{equation}
r \equiv -\frac{\frac{\epsilon_1 \text{Ma}}{2}-\frac{\text{Bo}}{3\text{Ca}}}{\frac{\epsilon_2 \text{Ma}}{2}}.
\end{equation}

\noindent For $r<0$ the lower layer is the initially unstable one, while for $r>0$ it is the upper layer that triggers the evolution of both films. If $\left|r\right|>1$ the system is unstable and $\Phi$ is either $0$ (in-phase) or $\pi$ (antiphase). If $r \in [-1,1]$, the system is stable and the phase between the two layers is given by $\Phi=\text{arctan}\left(-\sqrt{1-r^2}/r\right)$, where the inverse tangent function gives a value within $[-\pi,0]$. These results are summarized in Fig.~\ref{im:summary}. The characterization of the triple-layer system requires both $r$ and $s$, as the $\omega_{\pm}$ eigenvalues are solely functions of $s$ whereas the eigenvectors can be expressed only in terms of $r$.

To support the findings from Eq.~\eqref{eq:ident_eigvec}, a second series of numerical simulations was conducted. The value of $r$ was varied between $[-2,2]$. Once again, silicone oil as the liquid medium (Table \ref{tab:mat_prop}) and air as the gaseous medium were used. The thicknesses of the air and liquid layers were $120 \mu \text{m}$. The variation of $r$ was achieved by varying the temperature difference between $3.25\text{K}$ and $19.60\text{K}$. The fourth column of Table \ref{tab:dim_quant} summarizes the dimensionless groups of this simulation. 

For $|r|>1$ the numerical parameters of the simulations and the corresponding system behavior were similar to the previous simulations. On the other hand, for $|r|<1$ no patterns emerge. Consequently, it is not possible to calculate a characteristic wavenumber $q_{char}$ in order to scale the simulated region. However, the long-wavelength approximation remains valid, and the actual values of the scaling parameters do not have any qualitative effect on the system behavior, as they do not change the essential physics. Therefore, lacking natural scaling quantities, for the purpose of numerical simulation, a scaling length and velocity can be chosen arbitrarily. To this end, $L=10h_{0,1}$ was used again as the lateral scaling length and $v_c=10^{-5}\text{m/s}$ as the scaling velocity. In this case a square with a side length of $150$ was used as the simulation domain, while the number of cells remained at $120\times120$. Instead of $2.5/\omega_{char}$, the simulated time period was $-250/s$ ($s<0$), since $\omega_{char}$ is also not defined here. As before, the initial layer thicknesses were modulated by a white-noise perturbation with an amplitude of $5\cdot10^{-2}$.

The evolution of the system was calculated with two independent methods, starting from the same initial conditions. In the full numerical approach, Eqs.~\eqref{eq:h_mom_eq_double_1}~and~\eqref{eq:h_mom_eq_double_2} were directly solved. For comparison, in the semi-analytical method the Fourier transforms of  the initial liquid patterns were computed and all Fourier components were evolved independently from each other by using Eq.~\eqref{eq:general_fourier_sol}. In this case, the growth or decay of modes is governed by Eq.~\eqref{eq:ident_eigval}. 
To this end, $(\widetilde{\Delta H}_{1+}(k_x,k_y),$$ \widetilde{\Delta H}_{2+}(k_x,k_y))$ and $(\widetilde{\Delta H}_{1-}(k_x,k_y),$$ \widetilde{\Delta H}_{2-}(k_x,k_y))$ at $\hat{\tau}=0$ were obtained through the white-noise initial condition. Subsequently, the position space representation of the patterns was calculated with the inverse Fourier formula of Eq.~\eqref{eq:fourier_form} for every time step. The relative difference between the numerical (${\Delta H_1}^n,{\Delta H_2}^n$) and semi-analytical (${\Delta H_1}^a,{\Delta H_2}^a$) results is defined by
\begin{equation}
\begin{aligned}
D(\tau)=\frac{1}{2}&{\Bigg \{}\frac{\int{[{\Delta H_1}^{n}(\tau,x,y)-{\Delta H_1}^{a}(\tau,x,y)]^2dxdy}}{\int{[{\Delta H_1}^{n}(\tau,x,y)]^2dxdy}}+\\
&+\frac{\int{[{\Delta H_2}^{n}(\tau,x,y)-{\Delta H_2}^{a}(\tau,x,y)]^2dxdy}}{\int{[{\Delta H_2}^{n}(\tau,x,y)]^2dxdy}}{\Bigg \}}.
\end{aligned}
\end{equation}

\noindent Table \ref{tab:max_rel_diff} summarizes the maximal values of $D(\tau)$ over the whole simulated time interval. The full nonlinear numerical solution agrees well with the semi-analytical result obtained from linear theory. This table serves also as a verification that the integration time steps chosen in the numerical simulation are sufficiently small.

\begin{table}
  \centering
  \begin{ruledtabular}
  \begin{tabular}{r r M}
    \multicolumn{1}{c}{$r$} & \multicolumn{1}{c}{$s$} & \multicolumn{1}{c}{$max(D)$} \\
	\thickhline
	\rule{0pt}{2.5ex}
    $-2.000$ & $2.315$ \hspace{8mm} & $1.841 \cdot10^{-3}$\\
    $-1.500$ & $3.637\cdot 10^{-1}$  & $1.077\cdot10^{-3}$\\
    $-0.900$ & $-2.582 \cdot10^{-2}$  & $2.538\cdot10^{-5}$\\
   $-0.400$ & $-7.149 \cdot10^{-2}$  & $1.359\cdot10^{-7}$\\
    $0.000$ & $-6.249 \cdot10^{-2}$  & $1.007\cdot10^{-5}$\\
    $0.400$ & $-4.018 \cdot10^{-2}$  & $4.419\cdot10^{-7}$\\
    $0.900$ & $-6.794 \cdot 10^{-3}$  & $1.552\cdot10^{-5}$\\
    $1.500$ & $3.309 \cdot10^{-2}$  & $2.608\cdot10^{-4}$\\
    $2.000$ & $6.371 \cdot10^{-2}$  & $5.274\cdot10^{-4}$
    \end{tabular}%
    \end{ruledtabular}
    \caption{ Maximal relative difference between numerical and semi-analytical results.}
  \label{tab:max_rel_diff}
\end{table}

Three further series of numerical simulations were conducted. In the first one, the initial liquid layer thickness was $h_{0,1}=60\mu \text{m}$, while the temperature difference was varied between $[0.81\text{K},4.90\text{K}]$. In the second and third set of simulations the temperature difference between the substrates was fixed, and the liquid layer thicknesses were varied. These parameters were either $T_1-T_2=3\text{K}$ while $h_{0,1} \in [47\mu\text{m},115\mu\text{m}]$, or $T_1-T_2=5\text{K}$ while $h_{0,1}\in[60\mu\text{m},149\mu\text{m}]$. In all three cases the parameters were chosen in such a fashion that $r$ took the same values as listed in Table \ref{tab:max_rel_diff}. In every simulation the maximal value of the relative difference from the semi-analytical predictions was smaller than $10^{-3}$. This parametric study indicates that the results detailed in this paper are not specific to the particular choice of the initial film heights, but are in fact generic as long as the long-wavelength approximation holds. 

\subsection{Non-identical layers}

From an experimental viewpoint, exactly identical layers are impossible to achieve. To obtain a more general solution of practical relevance, in the following it is assumed that the initial thicknesses of both layers differ slightly, i.e., $\chi \neq 1$. As before, liquids with the same material properties (and hence, within the present formulation, with equal dimensionless groups)  are considered. In this case, 

\vspace{-12pt}
\begin{equation}
\begin{aligned}
\phi_1&=\epsilon_2\\
\phi_2 &= \epsilon_1+\delta \cdot \frac{\kappa D_0}{[\kappa\left(D_0-2-\delta\right)+2+\delta]^2} =\\
&= \epsilon_1+\delta \cdot (\epsilon_1+\epsilon_2),
\end{aligned}
\end{equation}

\noindent where $\delta=\chi-1=\left(h_{0,2}-h_{0,1}\right)/h_{0,1}$ is the difference between the initial liquid layer thicknesses. Reevaluation of the matrix $\boldsymbol{\mathrm{M}}$ of the linearized equations indicates the presence of additional terms of different orders of $\delta$, which make the analysis considerably more complicated, as if the film thicknesses were identical. Thus, as a simplification it is assumed that $\delta^2\ll1$ so that the governing equations can be linearized in $\delta$.  However, the linearization of $\epsilon_1$ and $\epsilon_2$ in terms of $\delta$ is unfavorable, as this would render the formulas less compact. Furthermore, leaving $\epsilon _1$
and $\epsilon_2$ in their original forms is not expected to significantly lower the accuracy. Therefore, to simplify the algebraic structure, linearization of these terms was not performed. The dimensionless time variable was again $\hat{\tau}=\tau/(3Ca)$. Apart from a linear correction in $\delta$, the eigenvectors of this system remain the same as in Eq.~$\eqref{eq:ident_eigvec}$. This correction does not result in any significant change in the system behavior. The eigenvalues are obtained as

\vspace{-12pt}
\begin{equation}\label{eq:non_ident_eigenvalue}
\begin{aligned}
\omega_{+/-}=&-q^4\left[1+\frac{3}{2}\delta\left(1\pm \frac{r}{\sqrt{r^2-1}}\right)\right]+\\
+&q^2\left[\pm\sqrt{s}-\frac{\delta}{2}\left(c\left(1\pm \frac{r}{\sqrt{r^2-1}}\right)\pm\frac{\sqrt{s}}{r^2-1}\right)\right],
\end{aligned}
\end{equation}

\noindent where for easier notation
\begin{equation}
c=\frac{3}{2}\text{Ca}\text{Ma}\left(-3\epsilon_2r +\epsilon_2-\epsilon_1\right)
\end{equation}

\noindent was introduced. Analyzing Eq.~\eqref{eq:non_ident_eigenvalue} one finds that, at $s>0$ and sufficiently small wavenumbers, the growth rate $\text{Re}(\omega_+)$ is still positive and the patterns remain locked in with respect to the horizontal coordinates since $\text{Im}(\omega_{\pm})=0$. In this regime the additional terms introduced by $\delta \neq 0$ only shift the characteristic wavenumber to different values. The oscillations, for which $\text{Im}(\omega_{\pm})\neq 0$, occur in the same region as for identical layers, i.e. for $s<0$, or equivalently, if $|r|<1$. In this regime the real part of $\omega_{\pm}$ reads

\vspace{-12pt}
\begin{equation}\label{real_nonident_osc}
\text{Re}(\omega_{\pm})=-q^4\left(1+\frac{3}{2}\delta\right)-q^2\frac{c \delta}{2}.
\end{equation}

\noindent The difference with the case of two identical layers is the following: if $c\delta<0$, then for sufficiently small wavenumbers the oscillatory deformation has a positive growth rate and instability occurs. Given the definition of $c$, the $c \delta<0$ condition can be reformulated to

\vspace{-12pt}
\begin{equation}\label{eq:instab_osci}
\begin{aligned}
r\gtrless r_0 &\text{ if } \delta\gtrless0,
\end{aligned}
\end{equation}

\noindent where $r_0=\left(\epsilon_2-\epsilon_1\right)/({3 \epsilon_2})$. Considering Eq.~\eqref{eq:epsilon_def} with $\kappa_1=\kappa_2$ (same liquid medium) and the relation $d_0>h_{0,1}+h_{0,2}$ one can show that at $H_1=H_2=1$ and for $\delta^2\ll1$ in fact $\epsilon_2-\epsilon_1<0$, thus $r_0<0$. The stability behavior of the triple-layer configuration is summarized in Table \ref{tab:stab_prop}. On the one hand, the sign of $r$ determines which liquid layer dominates the overall system dynamics. On the other hand, the value of $r$ relative to $r_0$ defines the stability of the oscillatory regime. 

\begin{table}
  \centering
  \begin{ruledtabular}
  \begin{tabular}{c N N N N }
   $r\in$ & $[-\infty,-1]$ & $[-1,r_0]$ & $[r_0,1]$ & $[1,\infty]$\\
    \thickhline
    $\delta<0$ & \rule{0pt}{2ex} stationary, unstable &  \rule{0pt}{2ex} oscillatory, unstable & oscillatory, stable &  stationary, unstable\\
    $\delta=0$ & stationary, unstable & oscillatory, stable & oscillatory, stable & stationary, unstable\\
    $\delta>0$ & stationary, unstable & oscillatory, stable & oscillatory, unstable & stationary, unstable
    \end{tabular}%
    \end{ruledtabular}
    \caption{ Stability regions of the system with two liquid layers. In stationary states the deformations are locked in with respect to the horizontal coordinates, whereas in oscillatory states they are mobile and move in-plane.}
  \label{tab:stab_prop}
\end{table}

The characteristic wavenumber of the oscillatory instability is $q_{char}=\sqrt{{-c \delta}/{(4+6\delta)}}$. Since $q_{char} \rightarrow 0$ with $r \rightarrow r_0$, this is a type-II-o instability \cite{cross_pattern}. Inserting $q_{char}$ into Eq.~\eqref{real_nonident_osc}, one finds the corresponding growth rate to be $\text{Re}(\omega_{char})={(c\delta)^2}/{(16+24\delta)}$. The angular velocity is $\text{Im}(\omega_{char})={c\delta}/{(4+6\delta)}\sqrt{-s}+O(\delta^2)$. The neutrally stable wavenumbers are $q_{min}=0$ and $q_{max}=\sqrt{2}\cdot q_{char}$, respectively. 

\begin{figure}
\centerline{\includegraphics[width=7.2cm]{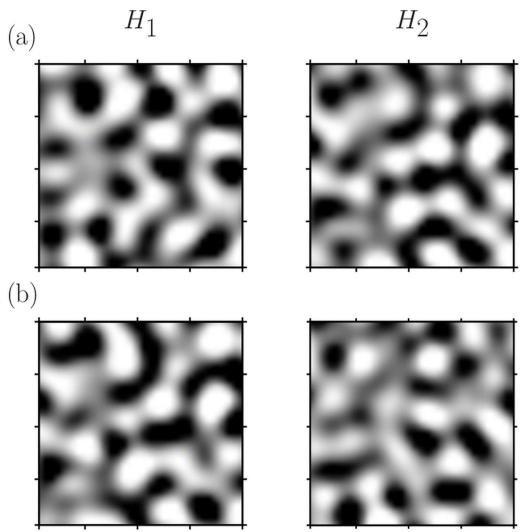}}
\caption{{The surface patterns $H_1$ and $H_2$ of $4\cdot \lambda_{char} \times 4\cdot \lambda_{char}$ sized regions extracted from the numerical simulations at (a)  $\hat{\tau}=0.998\cdot2.5/{\omega_{char}}$ and (b) $\hat{\tau}=2.5/{\omega_{char}}$.  The parameters are $s=-0.056$, $r=0.152$, and $c=-0.532$. The grayscale extends from $0.993$ to $1.007$.}\label{im:osci_pic}}
\end{figure}

For illustrative purposes the typical behavior of the oscillatory instability is shown in Fig.~\ref{im:osci_pic}. A $4\cdot \lambda_{char} \times 4\cdot\lambda_{char}$ sized rectangular cutout from the $15\cdot \lambda_{char} \times 15\cdot\lambda_{char}$ computational domain as obtained from numerical simulation is displayed. In this simulation the phase shift between the different layers was $\Phi\approx -1$. In the first row of Fig.~\ref{im:osci_pic}, the film height distributions of film 1 (left) and 2 (right) is shown at $\tau = 0.998 \cdot 2.5/\omega_{char}$. In the second row, the corresponding film height distributions at $\tau = 2.5/\omega_{char}$ are displayed. From the latter it is apparent, that in contrast to the spatially fixed patterns particularly observed in two-layer systems, the locations of the elevations change during the evolution of the oscillatory instability. Furthermore, Fig.~\ref{im:osci_pic} also illustrates that, contrary to the stationary instabilities, the phase shift between the two layers causes the patterns to differ from each other.

\begin{figure}
\includegraphics[width=\linewidth]{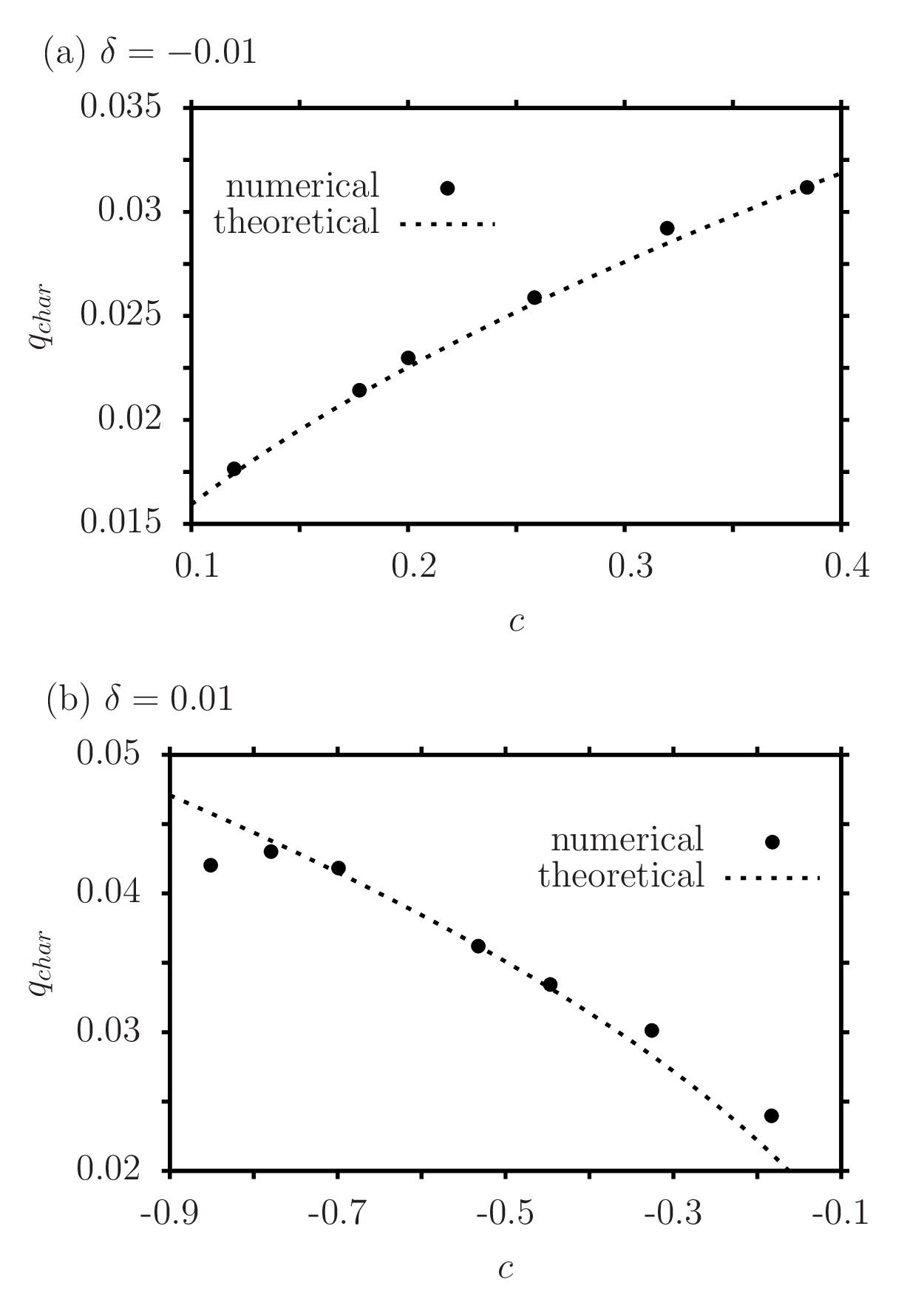}
\caption{The characteristic wavenumber of the oscillatory pattern emerging for (a) $\delta=-0.01$ and for (b) $\delta=0.01$.}\label{im:osc_char}
\end{figure}

For verification, the semi-analytical results were compared in detail with numerical simulations. As before silicone oil was considered as the liquid medium, and the thickness of the air layer was equal to the thickness of the lower liquid layer. This value was fixed at $h_{0,1}=120\mathrm{{\mu}m}$. For the first series of simulations, the initial thickness of the upper layer was set to $h_{0,2}=h_{0,1}\cdot 1.01$ ($\delta=0.01$). For the second series it was $h_{0,2}=h_{0,1}\cdot 0.99$ ($\delta=-0.01$). The $c$, $r$ and $s$ values were varied by changing the temperature difference between the substrates from $4.2\text{K}$ to $8.4\text{K}$. For both configurations, the simulations were conducted in the range of $r$ where the oscillatory instability occurs. The computational domains in space and time were defined in the same way as in the previous simulations, i.e., by using $q_{char}$ and $\text{Re}(\omega_{char})$. For both values of $\delta$ and near the $r\rightarrow r_0$ limit, the oscillation frequency becomes very high compared to the growth rate, i.e., $\text{Im}(\omega_{char})\gg\text{Re}(\omega_{char})$. Thus the computational costs to sufficiently resolve the emerging patterns in time also increases considerably. As a consequence, for $\delta = -0.01$ it was not possible to examine the whole range of $r \in [-1,r_0]=[-1,-0.61]$. Instead, the simulations were limited to $[-0.95,-0.72]$. In the simulated regime, $c\in[0.12,0.38]$ and $s\in[-5.3\cdot10^{-2},-1.3\cdot10^{-2}]$. Similarly, for the numerical simulations employing $\delta=0.01$, $r$ was varied between $[-0.36,0.9]$ instead of the full interval $[-0.59,1]$ and $c\in[-0.85,-0.18]$ while $s\in[-7.9\cdot10^{-2},-6.9\cdot10^{-3}]$. The results for the characteristic wavelengths of the patterns are depicted in Fig.~\ref{im:osc_char}. For $\delta=-0.01$ the maximal relative difference between the linear theory and the full numerical simulation was $2.6\%$, while for $\delta=0.01$ it was $12.9\%$. The latter is a significant deviation, which appeared at $r=-0.36$. This is the point closest to $r_0$. However, at other points the numerical and theoretical results agree reasonably well.
\\
\vspace{-12pt}

\section{Conclusion}

This article focuses on the effects of thermal coupling on the pattern evolution of two liquid layers placed opposite to each other and separated by a thin air layer. Specifically, the scenario was studied where the lower layer undergoes a gravity-stabilized long-wavelength B\'enard-Marangoni (BM) instability, while the upper one evolves under a thermocapillarity-stabilized long-wavelength Rayleigh-Taylor (RT) instability. For simplicity, the films are assumed to have similar initial thicknesses and the same material properties. The system was analyzed using linear stability theory as well as numerical solutions of the full nonlinear set of equations. A good overall agreement between the analytical and the numerical results was found. It was shown that the coupling can result in a qualitative difference in the evolution of the system compared to the conventional case where both instabilities evolve independently. Not only does the characteristic wavelength of the patterns increase, but the two layers may stabilize or destabilize each other. While without the coupling one of the layers would always be unstable while the other is stable, it was found that in the coupled system the stability behavior of the layers is synchronized. Moreover, for certain parameter ranges oscillatory instabilities were observed, which do not appear in the double-layer configuration. This is similar to the behavior found in coupled Turing pattern formations. It can be concluded that the coupling of two self-organizing systems can result in new modes of pattern formation not present in the individual systems.  

\begin{acknowledgments}
Funding by the German Research Foundation (DFG), Grant No. DI 1689/1-1, is gratefully acknowledged.
\end{acknowledgments}

\vspace{-12pt}
\appendix*
\section{Approximation of the characteristic velocity}\label{appendix_a}

The characteristic velocity $v_c$ is needed for the evaluation of $\text{Re}$, $\text{Ca}$, $\text{Ma}$ and $\text{Pe}$. To confirm that the approximations $a\cdot \text{Re}\ll1$ and $a\cdot \text{Pe}\ll1$ are valid during the whole process, an upper estimate of $v_c$ was used for the calculation of the dimensionless numbers.

The $x$-component of the dimensional in-plane velocity of the lower layer can be calculated from the momentum equations. At the interface, where the lateral velocities are the largest, the expression reads
\vspace{-12pt}

\begin{equation}\label{eq:v_x_nondim}
\begin{aligned}
v_x=&\left(\frac{\sigma}{\mu}\frac{\partial^3h_1}{\partial x^3}-\frac{g\rho}{\mu}\frac{\partial h_1}{\partial x}\right)\left(h_1\cdot h_{1,0}-\frac{h_{1,0}^2}{2}\right)+\\
+&\frac{\sigma_T(T_2-T_1)}{\mu \cdot h_{1,0}}h_{1,0}\left(\epsilon_1\frac{\partial h_1}{\partial x}+\epsilon_2\frac{\partial h_2}{\partial x}\right).
\end{aligned}
\end{equation}

\noindent The characteristic velocity was assumed to be the maximal value of this function. The calculation of $v_x$ requires expressions for the liquid thicknesses $h_1$ and $h_2$. In the present approximation, the surface deformation of the layers was assumed to be sinusoidal with a wavenumber of $q_{char}$. In the simulations the maximal relative deformations of the layers never exceeded $30\%$ of the initial film height. Thus, the amplitude of the sinusoid for the more unstable layer was set to $30\%$, whereas for the more stable one it was calculated with Eq.~\eqref{eq:ident_eigvec}. In summary, the assumed thickness functions were

\vspace{-12pt}
\begin{equation}\label{eq:ass_def}
\begin{aligned}
h_1&=h_{1,0}\cdot\left(1+A\cdot \text{sin}\left(2\pi  \frac{a\cdot x}{h_{1,0}}\right)\right)\\
h_2&=h_{2,0}\cdot\left(1+A\frac{\widetilde{\Delta H_2}}{\widetilde{\Delta H_1}}\cdot \text{sin}\left(2\pi  \frac{a\cdot x}{h_{1,0}}\right)\right),\\
\end{aligned}
\end{equation}

\noindent where $A$ is chosen in such a fashion that $max(A,A{\widetilde{\Delta H_2}}/{\widetilde{\Delta H_1}})$$=$$0.3$ and $a$ is obtained \nolinebreak from the linear analysis. The scaling velocities \nolinebreak for \nolinebreak the first set of simulations of Table \ref{tab:dim_quant}  were $v_c$$\in$$[6.33\cdot10^{-5}\text{m/s}$$,5.44\cdot10^{-4}\text{m/s}]$. For the second series they were approximated by $v_c$$\in$$[1.30\cdot10^{-5}\text{m/s},4.40\cdot10^{-4}\text{m/s}]$. As a consequence of the continuity of the velocities at the material interfaces \cite{merkt_liquid-liquid}, these results can be also used to approximate the characteristic lateral velocity of the air layer. 

\end{document}